\documentclass[12pt]{article}

\usepackage{a4}
\usepackage{epsfig}

\newlength{\abstwidth}
\newcommand{\mT}{m_{V,{\mathrm T}}}
\newcommand{\mL}{m_{V,{\mathrm L}}}
\newcommand{\mTrho}{m_{\rho,{\mathrm T}}}
\newcommand{\mLrho}{m_{\rho,{\mathrm L}}}
\newcommand{\mTphi}{m_{\phi,{\mathrm T}}}
\newcommand{\mLphi}{m_{\phi,{\mathrm L}}}
\renewcommand{\d}{{\mathrm d}}
\newcommand{\e}{{\mathrm e}}
\newcommand{\p}{{\mathrm p}}
\newcommand{\T}{{\mathrm T}}
\renewcommand{\L}{{\mathrm L}}
\newcommand{\JP}{{\mathrm J}/\psi}

\begin{document}

\def\lsim{\mathrel{\rlap{\lower4pt\hbox{\hskip1pt$\sim$}}
    \raise1pt\hbox{$<$}}}         
\def\gsim{\mathrel{\rlap{\lower4pt\hbox{\hskip1pt$\sim$}}
    \raise1pt\hbox{$>$}}}         

\pagestyle{empty}

\begin{flushright}
CERN--TH/98--294\\ 
hep-ph/9810370
\end{flushright}

\vspace{\fill}

\begin{center}
{\Large\bf Vector-Meson Electroproduction\\[1ex]
from Generalized Vector Dominance}\\[1.8ex]
{\bf Dieter Schildknecht${}^{a}$} \\[1.2mm] 
Theoretical Physics Division, CERN, 
CH-1211 Geneva 23 \\ 
and \\ 
Fakult\"{a}t f\"{u}r Physik, Universit\"{a}t Bielefeld, 
D-33501 Bielefeld \\[1.5ex]
{\bf Gerhard A.\ Schuler${}^{b}$} \\[1.2mm] 
Theoretical Physics Division, CERN, 
CH-1211 Geneva 23 \\[1.3ex]
and \\ 
{\bf Bernd Surrow} \\[1.2mm] 
Experimental Physics Division, CERN, 
CH-1211 Geneva 23 \\
\end{center}

\vspace{\fill}
\begin{center}
{\bf Abstract}\\[2ex]
\begin{minipage}{\abstwidth}
Including destructively interfering off-diagonal transitions of
diffraction-dis\-so\-ci\-a\-tion type, we arrive at a formulation of GVD for
exclusive vector-meson production in terms of a continuous spectral
representation of dipole form.
The transverse cross-section, 
$\sigma_{\T,\gamma^{*} \p \rightarrow V\p }$, behaves asymptotically as 
$1/Q^{4}$, while $R_{V}\equiv\sigma_{\L,\gamma^{*} \p \rightarrow V\p }%
/\sigma_{\T,\gamma^{*} \p \rightarrow V\p }$ becomes asymptotically constant. 
Contributions violating $s$-channel helicity conservation stay at the 
$10$--$15\%$ level established in low-energy photoproduction and diffractive 
hadron--hadron interactions. The data for $\phi$- and $\rho^{0}$-meson 
production for $0\lsim Q^{2} \lsim 20\,$GeV$^{2}$ from HERA are found to be in 
agreement with these predictions.
\end{minipage}
\end{center}

\vspace{\fill}
\noindent
CERN--TH/98--294\\
September 1998

\vspace{\fill}
\noindent
\rule{60mm}{0.4mm}

\vspace{0.1mm} 
\noindent
${}^a$ Supported by the BMBF, Bonn, Germany, Contract 05 7BI92P 
and the EC-network contract CHRX-CT94-0579.\\
${}^b$ Heisenberg Fellow; 
supported in part by the EU Fourth Framework Programme
``Training and Mobility of Researchers'', Network 
``Quantum Chromodynamics and the Deep Structure of Elementary Particles'', 
contract FMRX-CT98-0194 (DG 12-MIHT). 
\clearpage
\pagestyle{plain}
\setcounter{page}{1} 

The key role played by the vector mesons in the dynamics of hadron photoproduction
on nucleons,
at energies sufficiently above the vector-meson production thresholds, 
became clear in the late sixties and
early seventies. Indeed, the total photoproduction cross-section on protons, 
$\sigma_{\gamma \p}(W^{2})$, was found to be related to forward vector-meson photoproduction,
$\d\sigma^{0}/\d t|_{\gamma \p \rightarrow V\p }(W^{2})$, extrapolated to 
$t=0$ \cite{ref1}\footnote{A precision evaluation of (1) requires
a correction for the (small) ratio of real to imaginary forward scattering amplitudes to be inserted
in the right-hand side of (1).},
\begin{equation}
\sigma_{\gamma \p}(W^{2})=\sum_{V=\rho^{0},\omega,\phi,\JP}\sqrt{16\pi}
\sqrt{\frac{\alpha\pi}{\gamma^{2}_{V}}}
\left(\frac{\d\sigma^{0}}{\d t}|_{\gamma \p \rightarrow V\p }
(W^{2})\right)^{1/2}\;,
\label{firstsumrule}
\end{equation}
and to the total cross-sections for the scattering of transversely polarized vector mesons
on protons, $\sigma_{V\p }$, obtained \cite{ref2}
by applying the additive quark model for hadron--hadron interactions
\begin{equation}
\sigma_{\gamma \p}(W^{2})=\sum_{V=\rho^{0},\omega,\phi,\JP}\frac{\alpha\pi}{\gamma^{2}_{V}}\sigma_{V\p }(W^{2})\;.
\label{secondsumrule}
\end{equation}
The factor $\alpha\pi/\gamma^{2}_{V}$ in (1) and (2) denotes the strength of the coupling
of the (virtual) photon to the vector meson $V$, as measured in $\e^{+}\e^{-}$ annihilation by the 
integral over the vector-meson peak:
\begin{equation}
\frac{\alpha\pi}{\gamma^{2}_{V}}=
\frac{1}{4\pi^{2}\alpha}\sum_{F}
\int\sigma_{{ \e^{+}\e^{-}\rightarrow V\rightarrow F}}(s)\d s\;,  
\label{couplingone}
\end{equation}
or by the partial width of the vector meson:
\begin{equation}
\Gamma_{V\rightarrow \e^{+}\e^{-}}=\frac{\alpha^{2}m_{V}}{12(\gamma^{2}_{V}/4\pi)}\;.
\label{couplingtwo}
\end{equation}
The sum rules (1) and (2) are based on 

$i)$ the direct couplings of the vector mesons to the photon and on
 
$ii)$ subsequent strong-interaction diffractive scattering of the vector mesons on the proton.

Relations (1) and (2) accordingly provide the theoretical basis for applying concepts of 
strong-interaction
physics, such as Regge-pole phenomenology, to the interaction of the photon with nucleons. Compare
\cite{ref3} for a recent analysis of the experimental data for the total photoproduction
cross-section in terms of Regge phenomenology.

The sum rule (1) is an approximate one. The fractional contributions of the different vector
mesons to the total cross-section, $\sigma_{\gamma \p}$, were found 
to be \cite{ref4}\footnote{Compare also the review \cite{Leith}.}
\begin{equation}
r_{\rho}=0.65\; , \quad r_{\omega}=0.08\; , \quad r_{\phi}=0.05\; , 
\label{rvalues}
\end{equation}
adding up to approximately $78\,\%$ of the total cross-section. An additional contribution
of $r_{\JP}\simeq\,1$--$2\,\%$ has to be added for the $\JP$ vector meson. To saturate
the sum rule (1), the contributions of the leading vector mesons have to be 
supplemented by more massive contributions also coupled to the photon, as observed in $\e^{+}\e^{-}$
annihilation. From the point of view of generalized vector dominance (GVD)
\cite{ref4}, 
the sum rules (1) and (2) appear as an approximation that is reasonable for the $Q^{2}=0$
case of photoproduction, while breaking down with increasing space-like $Q^{2}$,
the role of $\rho^{0}$, $\omega$ and $\phi$ being taken over by more massive states.

Relations (1) and (2) implicitly contain the propagators of the different vector mesons. Being
evaluated for real photons at $Q^{2}=0$, no explicit propagator factors appear
in (1) and (2), and the photon vector-meson transition with subsequent vector-meson
propagation is reduced to a multiplication
of the various cross-sections by coupling constants characteristic of the vector-meson photon
junctions.
It was pointed out a long time ago 
\cite{ref5}\footnote{See also \cite{Cho}.}
that an experimental study of vector-meson
electroproduction would provide an additional and particularly 
significant test of the 
underlying photoproduction dynamics. 

The presence of the vector-meson propagators 
in the respective production amplitudes for the various vector mesons would be
explicitly tested in vector-meson electroproduction.
In addition, vector-meson production by virtual photons, at values
of $Q^2 \gg m_V^2$, would allow to test the expected dominance of the
production by longitudinal photons over the production by transverse ones.
Moreover, the hypothesis of helicity conservation with respect 
to the centre-of-mass frame of the reaction $\gamma^{*} \p \rightarrow V\p $,
the hypothesis of `$s$-channel helicity conservation' (SCHC), 
introduced in \cite{ref5} by generalizing experimental results from
photoproduction \cite{experi} to electroproduction, would
become subject to experimental tests.

More recently, it was conjectured \cite{ref6}--\cite{ref10} 
that vector-meson electroproduction
at large values of $Q^{2}$ was calculable in perturbative QCD (pQCD) 
and would provide experimental
tests of it. We will comment on the results from the pQCD approach below.

Expressing the cross-section for forward ($t\simeq 0$) production of vector mesons on nucleons
by transversely polarized virtual photons in terms of 
the respective real-photon cross-sections,
we have the vector-meson dominance model (VDM) prediction \cite{ref5}
\begin{equation}
\frac{\d\sigma^{0}_{\T}}{\d t}|_{\gamma^{*} \p \rightarrow V\p }(W^{2},Q^{2})=
\frac{m_{V}^{4}}{(Q^{2}+m_{V}^{2})^{2}}
\frac{\d\sigma^{0}}{\d t}|_{\gamma \p \rightarrow V\p }(W^{2})\;.
\label{VMDpredT}
\end{equation}
For longitudinally polarized virtual photons, as a consequence of the coupling of the vector
meson $V$ to a conserved source as required by electromagnetic current conservation, the
result \cite{ref5}
\begin{equation}
\frac{\d\sigma^{0}_{\L}}{\d t}|_{\gamma^{*} \p \rightarrow V\p }(W^{2},Q^{2})= 
\frac{m_{V}^{4}}{(Q^{2}+m_{V}^{2})^{2}}\xi_V^{2}\frac{Q^{2}}{m_{V}^{2}}
\frac{\d\sigma^{0}}{\d t}|_{\gamma \p \rightarrow V\p }(W^{2})
\label{VMDpredL}
\end{equation}
was obtained.
Both relations (6) and (7) contain 
SCHC. 
The parameter $\xi_V$ denotes the
ratio of the imaginary forward scattering amplitudes for the scattering of longitudinally
and transversely polarized vector mesons and may in principle depend on the vector meson $V$ under
consideration and on the energy $W$. The 
value of $\xi_V = 1$ corresponds to the conjecture of helicity
independence of vector-meson nucleon scattering in the high-energy limit.

The predictions (6) and (7) for vector-meson production by virtual photons are based on the 
idealization that the propagation of the single vector meson $V$ is responsible for the $Q^{2}$
dependence of the diffractive electroproduction of that vector meson $V$. This idealization
is by no means true in nature. Time-like photons also couple to the continuum of 
hadronic states beyond $\rho^{0}$, $\omega$, $\phi$, etc., resulting from $\e^{+}\e^{-}$ annihilation
into 
quark--antiquark pairs, and vector-meson forward scattering need not necessarily
be `diagonal' in the masses of the ingoing and outgoing vector mesons. The process of diffraction
dissociation, corresponding in the present context 
to `off-diagonal' transitions 
such as $\rho^{0}\p \rightarrow \rho^{'0}\p$ etc.,
is in fact well known to exist in hadron--hadron interactions, as explicitly observed 
in proton--proton scattering \cite{ref11}. 

The modification of the vector-meson electroproduction cross-section resulting from
the inclusion of off-diagonal transitions of the diffraction-dissociation type 
was investigated in \cite{ref12}.
For definiteness, in \cite{ref12}, the calculation of vector-meson production was based
on a spectrum of an infinite series of vector-meson states coupled to the photon in a manner
that assures duality to quark--antiquark production in 
$\e^{+}\e^{-}$ annihilation. Under the 
fairly general assumption of a power law for the diffraction-dissociation 
amplitudes (at zero $t$) in terms 
of the ratios of the masses of the diffractively produced vector states 
\begin{equation}
 T[V \p \to V_N\, \p] = c_0\,  T[V \p \to V\, \p]\, 
  \left( \frac{m_1}{m_N} \right)^{2p+1} \qquad (N=1,2,3,\ldots)
\ ,
\label{Tratio}
\end{equation}
an intuitively very simple and satisfactory result was obtained.

The sum of the poles in the transverse amplitude for 
$\gamma^{*} \p \rightarrow V\p $ was shown 
to sum up approximately to a single pole, the pole mass $m_{V}$ of the vector meson $V$ 
being changed, however, to a value of $\mT$
different from $m_{V}$. Prediction (6), taking into account off-diagonal transitions
as embodied in 
GVD, thus becomes \cite{ref12}\footnote{The simple result (\ref{fraasT})
is an approximation that coincides with the full GVD result at 
$Q^2=0$ and $Q^2 \to \infty$,
but may vary by $\sim 10\%$ at intermediate $Q^2$ values.} 
\begin{equation}
\frac{\d\sigma^{0}_{\T}}{\d t}|_{\gamma^{*} \p \rightarrow V\p }(W^{2},Q^{2})
=
\frac{\mT^{4}}{(Q^{2}+\mT^{2})^{2}}\, 
\frac{\d\sigma^{0}_{\T}}{\d t}|_{\gamma \p \rightarrow V\p }(W^2)\;.
\label{fraasT}
\end{equation}
For destructive interference among neighbouring vector-meson states, 
incorporated in \cite{ref12} through an alternating-sign series
of vector-meson states, one finds
\begin{equation}
\mT < m_{V}\;.
\label{mVTrel}
\end{equation}
The alternating-sign assumption was originally motivated 
by GVD investigations of the total virtual photo-absorption 
cross-section \cite{ref13}. (For a recent analysis see e.g.\ \cite{FGM}). 
Alternating signs appear also in a recent QCD analysis \cite{ref10} 
of $\rho$, $\rho'$, and $\rho''$ diffractive photo- and electroproduction.
The precise value of $\mT$ in (\ref{fraasT})
depends on the details of the strong amplitude, 
i.e.\ on the strength $c_0$ and the exponent $p$ of the 
power-law ansatz (\ref{Tratio}) for (spin-conserving) diffraction
dissociation. 

With (\ref{fraasT}), the asymptotic behaviour of the transverse 
forward-production cross-section in (off-diagonal) 
GVD becomes
\begin{equation}
\frac{\d\sigma_{\T}^{0}}{\d t}|_{\gamma^{*} \p \rightarrow V\p }(W^{2},Q^{2}
 \rightarrow \infty ) = \frac{\mT^{4}}{Q^{4}}
\frac{\d\sigma^{0}}{\d t}|_{\gamma \p \rightarrow V\p }(W^2)
\ .
\label{sigTasym}
\end{equation}
%
While the power of $Q^{2}$ in (\ref{fraasT}) and (\ref{sigTasym}) 
remains unchanged with respect to 
(6), the normalization of the asymptotic cross-section relative to photoproduction 
is affected by the fourth power of $\mT$. Concerning sum rule (1):
it is unaffected by the introduction of off-diagonal
terms, since the initial photon remains, when passing from the 
left-hand side to right-hand side of (1). 
In relation (2), off-diagonal terms with destructively interfering amplitudes
imply multiplication of each $\sigma_{V\p }$ 
by a specific correction factor somewhat 
smaller than unity \cite{ref12}.

The result (\ref{fraasT}) (or rather the underlying amplitude) 
with the constraint (\ref{mVTrel}) 
in \cite{ref12} was obtained
by straightforward summation of an alternating series. 
In view of the ensuing extension to 
longitudinal photons, we note that the transverse amplitude
may, to a good approximation, be represented by a sum of 
dipole terms\footnote{Although always possible, 
given the result (\ref{fraasT}) of the series, 
the dipole approximation of two neighbouring terms in the series is most
natural for the choice $p=0$ in (\ref{Tratio}), the value supported
by diffraction-dissociation data \cite{ref11}.} 
by combining neighbouring terms in the series. Switching 
to an equivalent continuum formulation, we obtain 
the following representation of the transverse amplitude as an integral over dipoles
\begin{equation}
A_{\T,\gamma^{*} \p \rightarrow V\p }(W^{2},Q^{2},t=0)=
\mT^{2}\int_{\mT^{2}}\frac{\d m^{2}}{(Q^{2}+m^{2})^{2}}
A_{\gamma \p \rightarrow V\p }(W^{2},t=0)
\ .
\label{ATcont}
\end{equation}
Note that the modified pole mass $\mT$ of the discrete formulation 
has turned into an effective threshold in (\ref{ATcont}).
Upon integration and squaring we immediately recover (\ref{fraasT}).

We note that our simple ansatz for diffraction dissociation does not
lead to the change of the $W$ dependence of vector-meson production 
with increasing $Q^{2}$ for which there is some
experimental indication \cite{Monteiro}. Such an effect can be 
incorporated into GVD by modifying the $W$ dependence of
diffraction dissociation. Any additional $W$-dependence in GVD 
is expected to enter via the ratio 
$x\simeq Q^{2}/W^{2}$ and yield an additional
(mild) $Q^{2}$ dependence beyond the propagator effect.

The impact of off-diagonal transitions on the result for longitudinally polarized 
virtual photons (7) was not explored in \cite{ref12}. 
The representation (\ref{ATcont}) for the 
transverse production amplitude as a continuous sum over dipole
contributions, abstracted from the assumed destructive interference 
between production amplitudes from 
neighbouring states, is well suited for a generalization to longitudinal
photons. Taking into account the coupling of the photon to a conserved source as transmitted
to the hadronic amplitude, we have
\begin{equation}
A_{\L,\gamma^{*} \p \rightarrow V\p }(W^{2},Q^{2},t=0)=
\xi_{V}\, \mL^{2}
\int_{\mL^{2}}\sqrt{\frac{Q^{2}}{m^{2}}}\frac{\d m^{2}}{(Q^{2}+m^{2})^{2}}
A_{\gamma \p \rightarrow V\p }(W^{2},t=0)\;.
\label{ALcont}
\end{equation}

In deriving (\ref{ALcont}), we have taken $\xi_V$ 
to be $m$-independent.
We expect the threshold mass of the longitudinal amplitude, $\mL$, 
to be larger than $m_{T}$, i.e. $\mT^{2}<\mL^{2}<m_{V}^{2}$. This is certainly true
if the occurrence of an additional inverse mass, associated with the 
extra $\sqrt{Q^2}$ factor in 
$A_{\L,\gamma^{*} \p \rightarrow V\p}$, 
is the only difference between the $m$-dependence of 
$A_{\T,\gamma^{*} \p \rightarrow V\p}$ and
$A_{\L,\gamma^{*} \p \rightarrow V\p}$. 
A priori, the transverse and longitudinal (strong-interaction) 
diffraction-dissociation amplitudes 
$T_{\T / \L}[V \p \rightarrow V_N \p]$ 
may possess different $m$-dependences 
($p_{\L} \neq p_{\T}$ in (\ref{Tratio})), 
thus affecting the ratio $\mL^2/\mT^2$. 

Integration of (\ref{ALcont}) yields
\begin{eqnarray}
\lefteqn{
A_{\L,\gamma^{*} \p \rightarrow V\p }(W^{2},Q^{2},t=0) 
= } 
\label{AL}
\\ & &
\xi_{V}\left[\frac{\pi}{2}\frac{\mL^{2}}{Q^{2}}
-\frac{\mL^{3}}{\sqrt{Q^{2}}(Q^{2}+\mL^{2})}
-\frac{\mL^{2}}{Q^{2}}\arctan\frac{\mL}{\sqrt{Q^{2}}}\right] 
\, 
A_{\gamma \p \rightarrow V\p }(W^{2},t=0) 
\nonumber\\ & & 
 \rightarrow  
\frac{2}{3}\xi_{V}\frac{\sqrt{Q^{2}}}{\mL} \, 
A_{\gamma \p \rightarrow V\p }(W^{2},t=0) 
\qquad {\rm for} \; Q^{2} \; \rightarrow 0 
\nonumber\\  & &
\rightarrow 
\frac{\pi}{2}\xi_{V}\frac{\mL^{2}}{Q^{2}} \, 
A_{\gamma \p \rightarrow V\p }(W^{2},t=0) 
\qquad {\rm for} \; Q^{2} \; \rightarrow \infty \; .
\nonumber
\end{eqnarray}

The above predictions for transverse and longitudinal 
production amplitudes are valid for high-energy 
($x = Q^2/(W^2+Q^2) \ll 1$) forward ($t\simeq 0$) production. 
It would be preferable to compare the predictions with forward-production
data, thus eliminating the influence of a possible $Q^{2}$ dependence 
of the slope of the $t$-distribution. 
No reliable data for forward production have been extracted from the 
experiments so far. Accordingly, in order to be able
to compare at all with data available at present, 
we ignore a possible $Q^{2}$ dependence of the $t$-distribution by
putting $b(0)/b(Q^{2})\simeq 1$, where $b$ is the slope parameter in 
the $t$-distribution, $\exp(-b|t|)$. From (\ref{fraasT}), 
the transverse production cross-section integrated over 
$t$
then becomes
\begin{equation}
\sigma_{\T,\gamma^{*} \p \rightarrow V\p }(W^{2},Q^{2})=
\frac{\mT^{4}}{(Q^{2}+\mT^{2})^{2}}\sigma_{\gamma \p \rightarrow V\p }(W^2)\;.
\label{sigT}
\end{equation}

A remark on SCHC is appropriate at this point. From photoproduction 
measurements at lower energies it is known \cite{Leith} that
SCHC is not strictly valid. It is violated (at non-zero $t$) at the
level of approximately $10\%$. 
In vector dominance this amount of helicity-flip contributions is
traced back to helicity-flip contributions in diffractive hadron
reactions which occur at approximately the same rate. It is natural,
accordingly, that the diffractive scattering of vector states leading
to (12) also violates SCHC at the level of $10\%$. Hence the ansatz (12)
has to hold as well for the helicity-flip transitions occurring
at finite $t$. The relative amount of helicity-flip contributions
has to remain at the level of $10\%$ found in photoproduction.


The same arguments on SCHC also hold for the
longitudinal cross section, or rather the 
longitudinal-to-transverse ratio $R_{V}$.  From (\ref{AL}) 
and (\ref{sigT}) we obtain
\begin{eqnarray}
\lefteqn{
R_{V}(W^2,Q^2)  =  \frac{\sigma_{\L,\gamma^{*} \p \rightarrow V\p }}%
{\sigma_{\T,\gamma^{*} \p \rightarrow V\p }} }
\label{RV}
\\ 
& = & \frac{(Q^{2}+\mT^{2})^{2}}{\mT^{4}} \xi_{V}^{2}\, 
\left[
  \frac{\pi}{2}\, \frac{\mL^{2}}{Q^{2}}
-\frac{\mL^{3}}{\sqrt{Q^{2}}(Q^{2}+\mL^{2})}
-\frac{\mL^{2}}{Q^{2}}\arctan\frac{\mL}{\sqrt{Q^{2}}}\right]^{2} 
\nonumber\\
& \rightarrow & 
\frac{4}{9}\xi_{V}^{2}\frac{Q^{2}}{\mL^{2}} 
\qquad\;\;\; {\rm for} \; Q^{2} \; \rightarrow 0 
\nonumber\\
& \rightarrow & 
\frac{\pi^{2}}{4}\xi_{V}^{2}\,
\frac{\mL^{4}}{\mT^{4}} \qquad {\rm for} \; Q^{2} \; \rightarrow \infty \; .
\nonumber
\end{eqnarray}
The approach to the large-$Q^2$ limit is rather slow, but 
note the enhancement factor $(\mL/\mT)^4$ in (\ref{RV}). 
For completeness, we quote also the total virtual-photon cross-section 
and its asymptotic limit
\begin{eqnarray}
\sigma_{\gamma^{*} \p \rightarrow V\p }(W^{2},Q^{2})
& \equiv &
\sigma_{\T,\gamma^{*} \p \rightarrow V\p } + \epsilon\, 
\sigma_{\L,\gamma^{*} \p \rightarrow V\p }
\nonumber\\ & = &
\sigma_{\T,\gamma^{*} \p \rightarrow V\p }\, 
\left(1+\epsilon\, R_{V}(W^{2},Q^{2})\right)
\label{sigtotal}
\\
& \to & \frac{\mT^4}{Q^4}\, \left( 1 + \epsilon\, \frac{\pi^2}{4}\, 
\xi_V^2\, \frac{\mL^4}{\mT^4} \right)\, \sigma_{\gamma \p \to V\p}(W^2)
\quad (Q^2 \to \infty)
\nonumber
\ .
\end{eqnarray}

In the comparison of our predictions with experiment, we proceed
in two steps. In a first step we consider the experimental evidence 
for the validity of SCHC, before we turn to a comparison of 
(\ref{sigT})--(\ref{sigtotal}) with 
HERA data\footnote{\label{footref}For the
preprint we consider only the statistical errors of the preliminary
data and postpone a systematic error analysis for the journal version
until the final experimental results become available.}.
The validity of SCHC is not only of much interest in itself, due
to the presence of the longitudinal degree of freedom of the virtual
photon in electroproduction, but is as well a
prerequisite for the determination of $R_V$, as long as data are
lacking for a direct separation of $\sigma_{\T,\gamma^*\p \to V\p}$
and $\sigma_{\L,\gamma^*\p \to V\p}$. 

A recent measurement by the ZEUS collaboration \cite{ZEUSprel}
of the full set of $15$ density matrix elements determing the vector-meson
($\rho^0$ and $\phi$) decay distribution \cite{Schilling} can be
analyzed in terms of helicity-conserving and helicity-flip amplitudes. 
Using parity invariance as well as natural-parity exchange, the number
of independent helicity amplitudes determing the density matrix
elements can be reduced to ten. This number is reduced to five, if
nucleon helicity-flip amplitudes are assumed to vanish. The normalized 
density matrix elements, accordingly, depend on four ratios of amplitudes, 
if we take the amplitudes to be purely imaginary. 
\begin{table}
\begin{center}
\begin{tabular}{|c|c||c|c|c|c|c|}
\hline
$W\,$[GeV] & $Q^2\,$[GeV$^2$] & 
$\frac{M(00)}{M(++)}$ & $\frac{M(+0)}{M(++)}$ & 
$\frac{M(+-)}{M(++)}$ & $\frac{M(0+)}{M(00)}$
 & $\chi^2/{\mathrm{d.o.f.}}$
\\ \hline
$9.4$ & $0$ & -- & $0.14\pm 0.02$ & -$0.05 \pm 0.02$ & -- & --
\\ \hline
$40$--$100$ & $3$--$5$ & $1.57$  & $0.081$ & $0.05$ & ~~$0.03~$ & $2.3~$ 
\\ \cline{2-7}
            & $5$--$30$ & $2.02$ & $0.24~$ & $0.01$ & - $0.03~$ & $0.41$
\\ \cline{2-7}
            & $3$--$30$ & $1.77$ & $0.15~$ & $0.04$ & - $0.001$ & $1.8~$
\\ \hline
\end{tabular}
\caption{The ratios of the helicity amplitudes 
$M(\lambda_\gamma,\lambda_\rho)$ for 
$\gamma^*\p \to \rho^0 \p$
obtained from a fit to the $\rho^0$ density matrix elements 
as measured at HERA \cite{ZEUSprel}.
Only the central results are quoted and 
the $\chi^2$ values are based on merely the statistical errors
(cf.\ footnote~\ref{footref}).
The photoproduction results (first row) are from \cite{Leith}. 
\label{tableone}}
\end{center}
\end{table}
In a fit to the $\rho^0$ density matrix elements,  
we have determined these ratios\footnote{Current
data are not yet precise enough to include
nucleon helicity-flip amplitudes in a (nine-parameter)
fit to the density-matrix elements. Nevertheless, such
a fit shows that our main conclusions remain unchanged:
The values for Rv remain consistent with the ones obtained
assuming SCHC, and (some) helicity-flip contributions 
are of the order of $15\%$.}.

As the third column in table~\ref{tableone} shows, the
production of longitudinal $\rho^0$ mesons by longitudinal photons 
strongly dominates the production of transverse $\rho^0$ mesons by 
transverse photons. The fourth column in table~\ref{tableone} 
shows that the helicity-flip amplitude for production of longitudinal 
$\rho^0$ mesons by transverse photons is suppressed to a value of
order $15\%$ relative to the (sub-)dominant transverse 
helicity-conserving amplitude. This result is consistent of what is 
known from photoproduction and hadron--hadron interactions at lower
energies \cite{Leith}. Finally, the last two columns in table~\ref{tableone}
show that the remaining helicity-flip amplitudes are small. 
Hence, the central predictions of vector dominance at large $Q^2$, 
the dominant role of longitudinal photons and helicity conservation to the
extent characteristic for diffractive hadron--hadron scattering, are
confirmed by the measurements. 

We note that a one-parameter fit to the data, assuming SCHC,
yields values of the longitudinal-to-transverse ratio 
$M(00)/M(++)$ consistent with the values from the four-parameter fit, the 
$\chi^2/{\mathrm{d.o.f.}} \simeq 2.8$, $2.0$, $3.2$ for the three $Q^2$ rows
of table~\ref{tableone}, respectively, 
being substantially worse, however. A small violation of SCHC 
is necessary indeed.

We finally note that the value for
$R_{\rho}$ obtained from table~\ref{tableone},
$R_{\rho} = \{ |M(00)|^2 + 2\, |M(0+)|^2 \} /%
 \{ |M(++)|^2 + |M(+0)|^2 + |M(+-)|^2 \} 
 \simeq |M(00)|^2/|M(++)|^2 
\simeq 2.4$ ($4.0$)
for $3 < Q^2 < 5\,$GeV$^2$ ($5 < Q^2 < 30\,$GeV$^2$) is
consistent with the previous determination of $R_{\rho}$ 
from ZEUS. For the previous determination, the results of which
will be shown below, the validity of SCHC had to be assumed.

We now turn to the $Q^2$ dependence and compare
predictions (\ref{sigT})--(\ref{sigtotal}) 
with experimental data from HERA \cite{ZEUSprel,HERAdata}
at an average $\gamma^{*}\p$ c.m.\ energy of 
$W=80\,$GeV ($50\,$GeV)  for $\phi$ ($\rho^0$) production\footnote{
At HERA energies, we may take the polarization parameter $\epsilon =1$.}. 
For a given vector meson $V$,
our predictions depend on four parameters, the two effective 
vector-meson masses $\mT$ and $\mL$, the ratio $\xi_V$ of the 
longitudinal-to-transverse strong-interaction amplitudes, and 
the photoproduction cross-section, i.e.\ (\ref{sigT}) at $Q^2=0$. 
The solid lines in Figs.~1--3 show the result of a simultaneous 
four-parameter fit
to the data for $\sigma_{\gamma^{*}\p \rightarrow V \p}$ and
$R_V$, performed separately for the $\rho^0$ and the $\phi$ meson.
The data are well described by the fits, with the parameters
\begin{eqnarray}
 \xi_{\rho} = 1.06\ , \quad & \mTrho^2 = 0.68\, m_{\rho}^2\ , 
 \quad &  \mLrho^2 = 0.71\, m_{\rho}^2\ , 
\label{fourfit}\\
          \xi_{\phi} = 0.90\ , 
 \quad &  \mTphi^2 = 0.43\, m_{\phi}^2\ , 
 \quad &  \mLphi^2 = 0.60\, m_{\phi}^2\ , 
\nonumber
\end{eqnarray}
and $\sigma_{\gamma \p \to \rho\p} = 11.1\,\mu$b, 
$\sigma_{\gamma \p \to \phi\p} = 1.2\,\mu$b. 
The statistical errors in the parameters are small compared with the
estimated systematic ones$^{\ref{footref}}$.

The quality of the fits strongly supports the underlying picture: 
the propagation of hadronic spin-1 states and 
destructive interference 
govern the $Q^2$ dependence of exclusive electroproduction 
of vector mesons at small $x$ and arbitrary $Q^2$. Both the asymptotic 
$1/Q^4$ behaviour of the cross section, see (\ref{sigtotal}), and
the flattening of $R_V$, see (\ref{RV}), are clearly visible in 
the data.
Moreover, the fitted values (\ref{fourfit}) are in accordance with theoretical 
expectation. The value of $\xi_V \simeq 1$, i.e.\ helicity independence
of diffractive vector-meson scattering, is very gratifying indeed.
The mass parameters, $\mT$ and $\mL$, show the theoretically 
expected ordering $\mT^2 < \mL^2 < m_V^2$. 

The values of $R_V$ obtained in the fit seem somewhat
low with respect to the central values of the data at large $Q^2$. 
This is of course merely a consequence of the fact that the large-$Q^2$
$R_V$ data hardly contribute to the overall 
$\chi^2$, owing to their large errors. 
Varying the four fit-parameters within one standard deviation from their
best-fit values, we find that a considerable spread in $R_V$ is allowed. 
In other words, with current data a precision determination of
our parameters is not yet possible. In fact, a two-parameter fit
results in a similar $\chi^2$ (dashed lines in Figs.~1--3) as 
the four-parameter fit. In the two-parameter fit, obtained by fixing 
$\xi_V=1$ and $\mL^2 = 1.5\, \mT^2$
(corresponding to an asymptotic value $R_V \to 5.5$), we find
\begin{equation}
  \mTrho^2 = 0.62\, m_\rho^2 \ , \qquad
  \mTphi^2 = 0.40\, m_\phi^2 \ , 
\label{twofit}
\end{equation}
and $\sigma_{\gamma \p \to \rho\p} = 11\,\mu$b, 
$\sigma_{\gamma \p \to \phi\p} = 1.0\,\mu$b. 
With respect to the results of the fits given in (18) and (19), 
it may be worth quoting 
the estimate $0.41\, m_{V}^{2} \lsim \mT^{2} \lsim 0.74\, m_{V}^{2}$ 
from \cite{ref12}, based on a reasonable choice of the 
diffraction-dissociation parameters in (8).

In Figs. 2b and 3b, we show the transverse cross-section, 
$\sigma_{\T,\gamma^{*}\p\rightarrow V\p}$. 
The data in Figs. 2b and 3b were extracted from the data
on $\sigma_{\gamma^{*}\p\rightarrow V\p}$ in Fig. 1 with the help
of our two-parameter fit\footnote{No other procedure to extract 
$\sigma_{\T,\gamma^{*}\p\rightarrow V\p}$ suggests itself, 
as the number of data points for $R_{V}$  is very small, 
and the $Q^{2}$ values for $\sigma_{\gamma^{*}\p\rightarrow V\p}$
and $R_{V}$ are not identical.} 
for $R_{V}$. Figures 2b and 3b demonstrate the
dramatic difference at large $Q^{2}$ between the data and the 
GVD prediction (15) with $\mT<m_{V}$ on the one hand, and the VMD prediction
(6), or rather (15) with $\mT\equiv m_{V}$, on the other hand.

Comparing the dotted VMD predictions in Figs.\ 2b and 3b 
for the transverse cross-section
$\sigma_{\T,\gamma^{*}\p\rightarrow V\p}$ with the data for
$\sigma_{\gamma^{*}\p\rightarrow V\p}$ in Figs.\ 1a and 1b, one
notices that the dotted curves would approximately describe the data for
$\sigma_{\gamma^{*}\p\rightarrow V\p}=
\sigma_{T,\gamma^{*}\p\rightarrow V\p}+
\sigma_{L,\gamma^{*}\p\rightarrow V\p}$. 
This, at first sight paradoxical, coincidence
of fits of $\sigma_{\T,\gamma^{*}\p\rightarrow V\p}+
\sigma_{\L,\gamma^{*}\p\rightarrow V\p}$, entirely 
based on the transverse VMD formula, was in fact
observed previously \cite{Monteiro,ref14,Erdmann} in fits that vary
the power of $(Q^{2}+m_{V}^{2})$ at fixed mass $m_{V}$. Implicitly the fits
obviously assume $\sigma_{\L,\gamma^{*}\p\rightarrow V\p}=0$, and, 
disregarding the information from vector-meson decay indeed seem to
confirm $\sigma_{\L,\gamma^{*}\p\rightarrow V\p}=0$. 
This conclusion is inconsistent, however, with the results of the above 
analysis of the $\rho^0$ density matrix elements. This analysis
establishes beyond any doubt that longitudinal $\rho^0$ mesons are
almost exclusively produced by  longitudinal (virtual) photons 
(compare table~\ref{tableone}). 
The mentioned approximate coincidence of fits based on the 
VMD formula for $\sigma_{\T,\gamma^{*}\p\rightarrow V\p}$ with the
data for $\sigma_{\T,\gamma^{*}\p\rightarrow V\p}+
\sigma_{\L,\gamma^{*}\p\rightarrow V\p}$ 
appears as a numerical accident.

Recent theoretical work on the electroproduction of vector mesons has been concentrated on attempts
to deduce the cross-sections from perturbative \cite{ref6}--\cite{IK} and 
non-perturbative \cite{ref10} QCD. 
For the production cross-section by transversely 
polarized vector mesons, the calculations typically lead to a 
strong asymptotic decrease, as $1/Q^{8}$, modified sometimes by 
additional corrections to become $1/Q^{7}$. It may be argued
\cite{ref8}
that the region of 
$Q^{2} \lsim 30\,$GeV$^{2}$ explored at present, in which experiments
find a fall-off rather like $1/Q^{4}$, is not 
sufficiently asymptotic for 
pQCD
to yield reliable results. Further experiments at still 
larger values of $Q^{2}$ will 
clarify the issue. 


As for the longitudinal-to-transverse ratio, $R_{V}$, 
pQCD
calculations led to the same result of a linear rise in $Q^{2}$ as the
simple VDM predictions, compare (6) and (7). Such a linear rise is 
always obtained, if electromagnetic current conservation is the only
source of the $Q^{2}$ dependence of $R_V$. For large $Q^{2}$, this linear
rise is in conflict with experimental results. A behaviour of the cross-section
for $Q^{2} \gg m^{2}_{\rho}$, for both the production of longitudinally
as well as transversely polarized $\rho^{0}$ mesons, somewhat closer 
to the experimental data, was obtained in \cite{ref9}; the calculation was 
based on open $q\bar{q}$ production and parton-hadron duality.
It is interesting to note that the resulting cross-sections have a VDM
form\footnote{Compare (37) and (38) in \cite{ref9}.} 
multiplied by correction factors depending on the scaling variable $x$.
The asymptotic form for $R_{V}$ derived in \cite{ref9}
has recently been reproduced in a calculation based on 
$\rho^0$- meson wave-functions \cite{IK}. In \cite{IK}, also
pQCD calculations of the helicity-flip amplitudes have been presented.
While the general trend of the helicity-flip amplitudes is correctly
reproduced, a detailed comparison shows that the $\chi^{2}$ of these
predictions is $\chi^{2}\simeq 44$, i.e.\ $\chi^2$ is not better than for 
a representation of the density matrix elements under the assumption
of SCHC (with a value of $\chi^2 \simeq 45$) as given above. 
The coincidence of the relative magnitude of the helicity-flip 
amplitudes at large $Q^{2}$ with 
the helicity-flip amplitudes in photoproduction and 
diffractive hadron physics remains unexplained in the pQCD approach. 


In summary, we have investigated electroproduction of vector mesons 
in GVD. We have shown that destructive interference between
neighbouring vector states naturally leads to the spectral representations
(\ref{ATcont}) and (\ref{ALcont}) of the (zero-$t$) amplitudes for
$\gamma^{*}_{\T,\L} + \p \rightarrow V_{\T,\L} + \p$. Both predictions, the
asymptotic $1/Q^2$ fall-off of the transverse amplitude  
and the approach of $R_V$ towards a constant value, 
are in good agreement with the experimental data. 
The expected hierarchy, $\mT^2 < \mL^2 < m_V^2$,  of the pole masses
$\mT$, $\mL$
and the helicity independence of the strong-interaction
amplitudes (reflected in $\xi_V \simeq 1$) strongly support the GVD
picture: the propagation of hadronic vector states determines, 
for arbitrary $Q^2$,  
the $Q^2$ dependence of vector-meson production by virtual photons
in the diffraction region of $x \simeq Q^2/W^2 \ll 1$.
Moreover as expected in this picture, 
SCHC is experimentally violated at the order of magnitude of $10\%$,
the value typical for diffractive hadron-hadron scattering and 
photoproduction.


Returning to our starting point, the photoproduction sum rules (1) and (2), 
the present analysis strengthens their dynamical content, which is to reduce
photoproduction to vector-meson-induced reactions. 
More generally, in conjunction with the experimental
observation of states with masses up to about $20\,$GeV \cite{ref15} in
diffractive production in DIS at small $x$ and up to large $Q^2$, 
the present investigation supports the point of view \cite{ref16} that 
propagation and diffractive scattering of hadronic vector states is the 
basic dynamical mechanism in DIS at small values of the scaling variable.

\noindent\\[2ex]
{\it Acknowledgement}\hfill\\
It is a pleasure to thank Teresa Monteiro and G\"{u}nter Wolf for 
useful discussions on the HERA data. 

\clearpage

\begin{figure}[ht]
\vspace*{-1.5cm}
\begin{center}
\noindent\epsfig{file=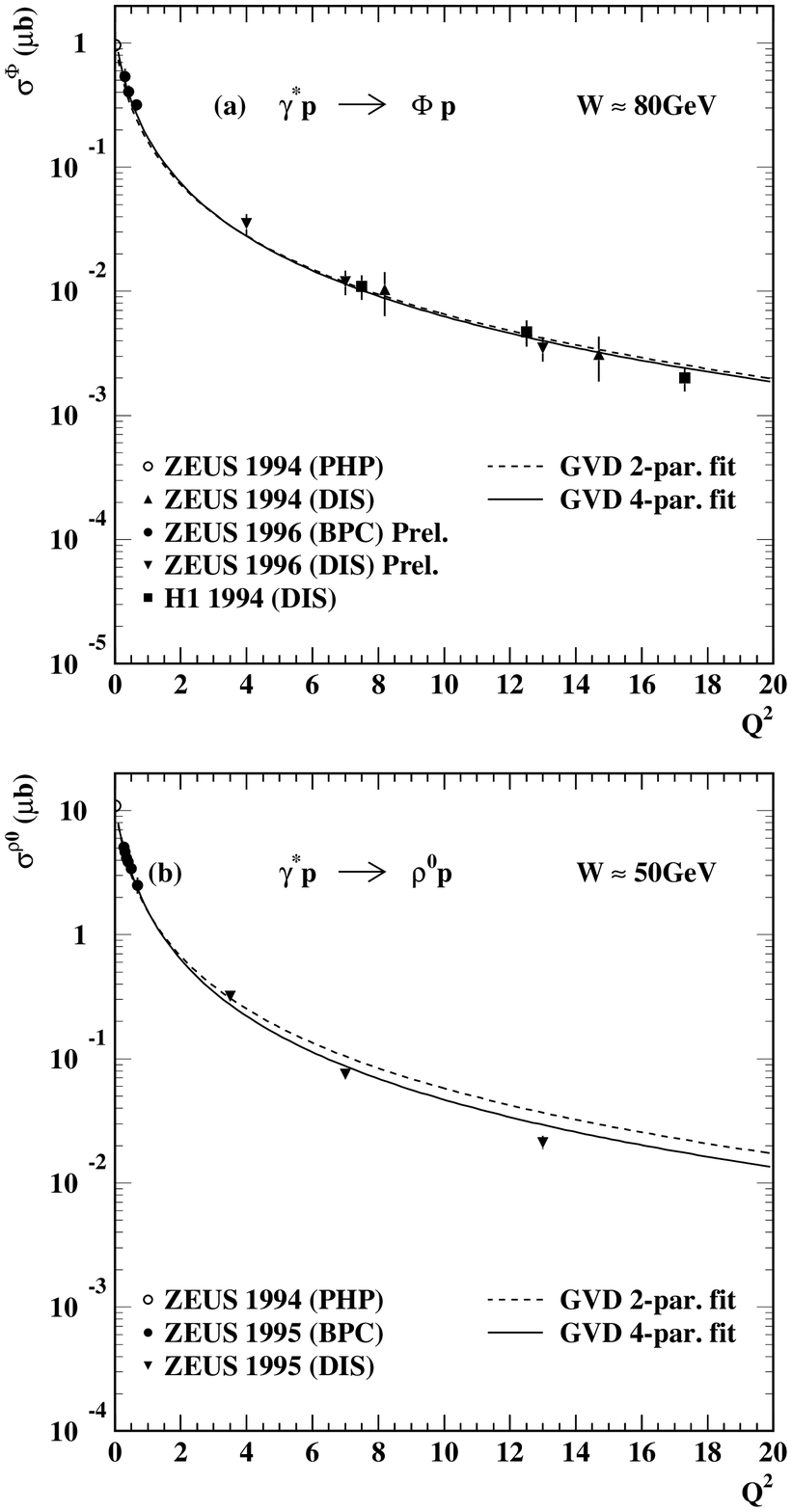,height=0.90\textheight}
\label{fig1}
\caption{Data for $\sigma_{\gamma^{*} \p \rightarrow \phi \p}$ (in (a)) and
for $\sigma_{\gamma^{*} \p \rightarrow \rho \p}$ (in (b)) from HERA 
compared with the GVD prediction (17). Solid lines: Four-parameter
fit with the values (18) of the fit parameters. Dashed line: Two-parameter
fit with the values (19) of the fit parameters.}
\end{center}
\end{figure}

\pagebreak

\begin{figure}[ht]
\begin{center}
\vspace*{-1.5cm}
\noindent\epsfig{file=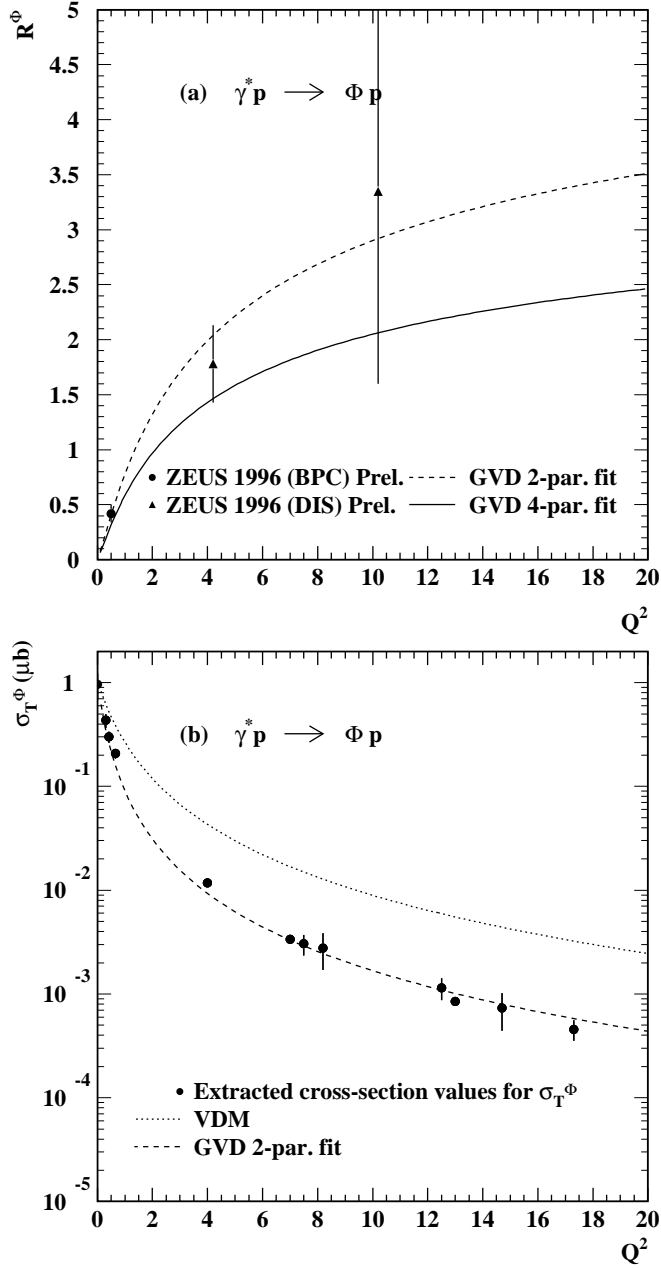,height=0.80\textheight}
\label{fig2}
\caption[]{(a) 
HERA data for the longitudinal-to-transverse ratio $R_{\phi}$ for 
$\phi$ production extracted from the
$\phi$ decay distribution using SCHC, compared with the GVD prediction (\ref{RV}). 
Solid line: Four-parameter
fit with the values (18) of the fit parameters. Dashed line: Two-parameter
fit with the values (19) of the fit parameters.\\
(b) Data for $\phi$ production by transversely polarized photons,
$\sigma_{\T,\gamma^{*} \p \rightarrow \phi \p}$, extracted 
from the measured values
of $\sigma_{\gamma^{*} \p \rightarrow \phi \p}$ by using the 
two-parameter $R_{\phi}$ fit shown in (a). Dashed line: 
GVD prediction (\ref{sigT}) with the two-parameter fit values
(19). Dotted line: VDM prediction, i.e.\ 
(\ref{sigT}) with $\mTphi \equiv m_\phi$.} 
\end{center}
\end{figure}

\pagebreak

\begin{figure}[ht]
\begin{center}
\vspace*{-1.5cm}
\noindent\epsfig{file=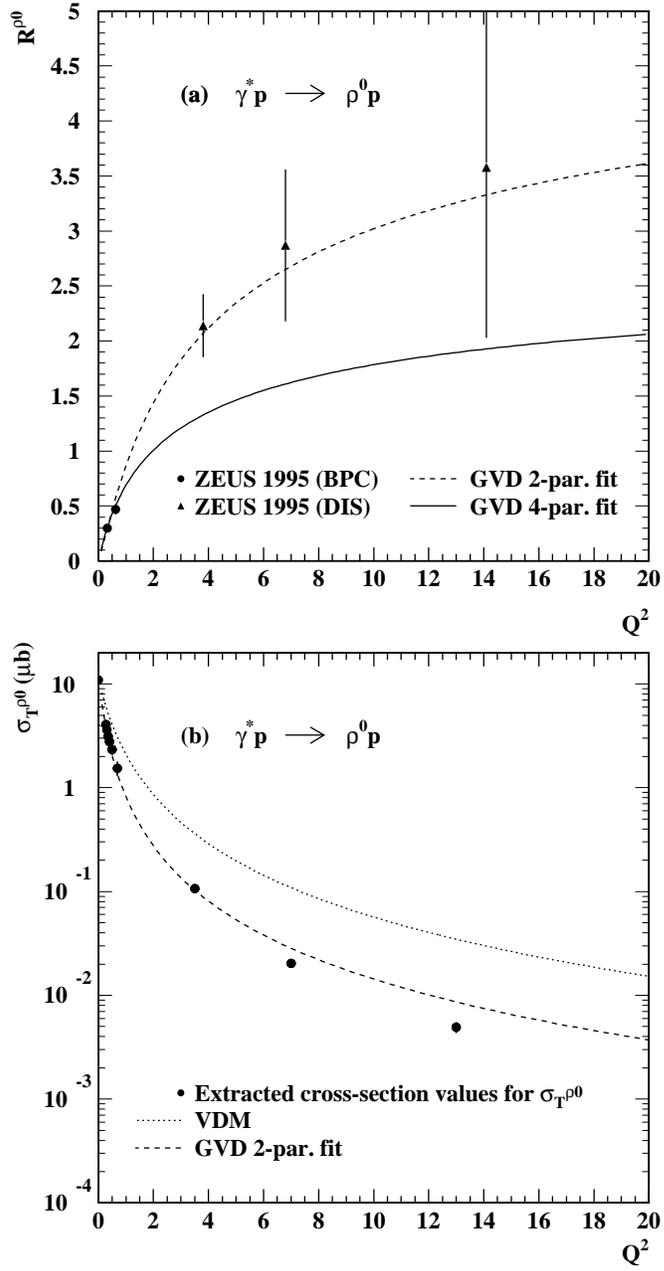,height=0.80\textheight}
\label{fig3}
\caption{As Fig.~2, but for $\rho^{0}$-meson production.}
\end{center}
\end{figure}

\end{document}